\documentclass[letterpaper, 10 pt, conference]{ieeeconf}  

\IEEEoverridecommandlockouts                              
\usepackage{graphics} 
\usepackage{epsfig} 
\usepackage{mathptmx} 
\usepackage{times} 
\usepackage{amsmath} 
\usepackage{amssymb}  

\usepackage{booktabs}

\usepackage{subcaption}
\usepackage{bm} 

\let\labelindent\relax
\usepackage[inline]{enumitem}   

\usepackage{caption}
\usepackage{multirow}

\usepackage[table]{xcolor}
\newcommand{\bestres}{green!15}

\usepackage{float}

\newtheorem{remark}{Remark}
\newtheorem{example}{Example}
\newtheorem{theorem}{Theorem}
\newtheorem{definition}{Definition}
\newtheorem{proposition}{Proposition}
\newtheorem{corollary}{Corollary}
\DeclareMathAlphabet{\mathcal}{OMS}{cmsy}{m}{n} 
\usepackage{hyperref}   
\newcommand{\aref}[1]{\hyperref[#1]{App.~\ref*{#1}}}

\makeatletter
\def\endthebibliography{%
  \def\@noitemerr{\@latex@warning{Empty `thebibliography' environment}}%
  \endlist
}

\usepackage{balance}    
\newcommand{\R}{\mathbb{R}}
\newcommand{\N}{\mathbb{N}}

\newcommand{\params}{\theta}                    
\newcommand{\dimParams}{{d_\theta}}             
\newcommand{\trueOp}{\mathbf{S}}            
\newcommand{\trueSys}{\mathbf{G}}           
\newcommand{\noiseModel}{\mathbf{H}}        
\newcommand{\trueSysFunc}[1]{G_{#1}}            
\newcommand{\approxSysFunc}[1]{\hat{G}_{#1}}    
\newcommand{\cont}{\mathbf{K}}              
\newcommand{\contFunc}[1]{K_{#1}}               
\newcommand{\approxOp}{\hat{\trueOp}}           
\newcommand{\approxOpFunc}[1]{\hat{S}_{#1}}     
\newcommand{\approxSys}{\hat{\trueSys}}         

\newcommand{\trueIn}[1]{u_{#1}}                 
\newcommand{\trueInMat}{\mathbf{u}}             
\newcommand{\trueInData}{\{\trueIn{\horizon-1:0}^\counter\}_{\counter=0}^{\numSamples-1}}
\newcommand{\trueNoisyOut}[1]{y_{#1}}

\newcommand{\trueNoisyOutMat}{\mathbf{y}}
\newcommand{\trueNoisyOutData}{\{\trueNoisyOut{\horizon-1:0}^\counter\}_{\counter=0}^{\numSamples-1}}
\newcommand{\exc}[1]{r_{#1}}
\newcommand{\excMat}{\mathbf{r}}

\newcommand{\noise}[1]{v_{#1}}
\newcommand{\noiseMat}{\mathbf{v}}
\newcommand{\whiteNoiseMat}{\mathbf{e}}
\newcommand{\trueState}[1]{x_{#1}}
\newcommand{\trueStateMat}{\mathbf{x}}
\newcommand{\approxIn}[1]{\hat{u}_{#1}}
\newcommand{\approxInMat}{\hat{\mathbf{u}}}
\newcommand{\approxOut}[1]{\hat{y}^\circ_{#1}} 
\newcommand{\approxOutMat}{\hat{\mathbf{y}}^\circ}
\newcommand{\approxOutData}{\{\hat{y}_{\horizon-1:0}^{\circ^\counter}\}_{\counter=0}^{\numSamples-1}}

\newcommand{\approxNoisyOutMat}{\hat{\mathbf{y}}}

\newcommand{\whatMat}{\hat{\bm{\omega}}}

\newcommand{\approxNoiseMat}{\mathbf{v}}

\newcommand{\dsFunc}{F}
\newcommand{\outputFunc}{H}

\newcommand{\upOp}{\bm{\Upsilon}}
\newcommand{\upOpKnot}{\bm{\Upsilon^o}}
\newcommand{\upFunc}[1]{\Upsilon_{#1}}
\newcommand{\upKnotFunc}[1]{\Upsilon^o_{#1}}
\newcommand{\eyeOp}{\mathbf{I}}
\newcommand{\aMat}{\mathbf{a}}
\newcommand{\bMat}{\mathbf{b}}
\newcommand{\aVec}[1]{a_{#1}}
\newcommand{\bVec}[1]{b_{#1}}

\newcommand{\dimIn}{{d_u}}
\newcommand{\dimOut}{{d_y}}
\newcommand{\dimState}{{d_x}}

\newcommand{\counter}{n}
\newcommand{\numSamples}{N}
\newcommand{\horizon}{T}
\newcommand{\dataset}{\mathcal{D}}

\newcommand{\eye}{I}
\newcommand{\zeroMat}{0}

\DeclareMathOperator*{\argmin}{arg\,min}

\renewcommand{\aref}[1]{\hyperref[#1]{App.~\ref*{#1}}}

\newcommand{\lp}{\ell_p}
\newcommand{\Lp}{\mathcal{L}_p}
\newcommand{\Ltwo}{\mathcal{L}_2}

\newcommand{\dirIDSup}{dir-ID}          

\newcommand{\dirICISup}{dir-ICI}        

\newcommand{\indirICISup}{indir-ICI}        

\newcommand{\cost}{J}

\definecolor{darkgray}{HTML}{314354}

\newcommand{\strat}[1]{\textcolor{darkgray}{Strategy \MakeUppercase{#1}}}

\title{\LARGE \bf
Neural Identification of Feedback-Stabilized Nonlinear Systems
}

\author{
    Mahrokh G. Boroujeni,\textsuperscript{1*} 
    Laura Meroi,\textsuperscript{1*}
    Leonardo Massai,\textsuperscript{1} 
    Clara L. Galimberti,\textsuperscript{2} \\
    and Giancarlo Ferrari-Trecate\textsuperscript{1}
    \thanks{
        \textsuperscript{1} Institute of Mechanical Engineering, EPFL, Switzerland.
        \texttt{\{mahrokh.ghoddousiboroujeni, laura.meroi, l.massai, giancarlo.ferraritrecate\}@epfl.ch}.
    }
    \thanks{
        \textsuperscript{2} Dalle Molle Institute for Artificial Intelligence (IDSIA), SUPSI, Switzerland.
        \texttt{clara.galimberti@supsi.ch}.
    }
    \thanks{
        \textsuperscript{*} 
        The first two authors collaborated equally.
    }
    \thanks{
        This work was supported as a part of NCCR Automation, a National Centre of Competence in Research, funded by the Swiss National Science Foundation (grant number 51NF40\_225155) and the NECON project (grant number 200021\-219431).
    } 
}

\begin{document}

\maketitle
\thispagestyle{empty}
\pagestyle{empty}

\begin{abstract}
Neural networks have demonstrated remarkable success in modeling nonlinear dynamical systems.  However, identifying these systems from closed-loop experimental data remains a challenge due to the correlations induced by the feedback loop. Traditional nonlinear closed-loop system identification methods struggle with reliance on precise noise models, robustness to data variations, or computational feasibility. 
Additionally, it is essential to ensure that the identified model is stabilized by the same controller used during data collection, ensuring alignment with the true system’s closed-loop behavior.
The dual Youla parameterization provides a promising solution for linear systems, offering statistical guarantees and closed-loop stability. However, extending this approach to nonlinear systems presents additional complexities.
In this work, we propose a computationally tractable framework for identifying complex, potentially unstable systems while ensuring closed-loop stability using a complete parameterization of systems stabilized by a given controller.
We establish asymptotic consistency in the linear case and validate our method through numerical comparisons, demonstrating superior accuracy over direct identification baselines and compatibility with the true system in stability properties.
\end{abstract}

\section{Introduction}\label{sec:intro}
Neural networks have shown strong performance in learning nonlinear dynamical systems from input-output data, due to their flexibility and expressive power~\cite{Sjoberg99NN, Pillonetto25NN}. However, in many practical situations, collecting data under open-loop conditions is impractical or unsafe, particularly for unstable systems. Even for stable systems, feedback control may be necessary for safety, efficiency, or operation near desired setpoints. For instance, drones are controlled to fly safely, robots are operated to achieve certain tasks, and biological systems are regulated by natural mechanisms. In many cases, disconnecting the controller to collect open-loop data is either infeasible or prohibitively costly.

In such scenarios, data is typically gathered while the system operates in closed-loop with a stabilizing controller~\cite{Vandenhof98CLissues, Forssell99Revisited, Maruta21ASF}. This motivates closed-loop system identification, which aims to learn models from such data. Compared to the open-loop case, this setting introduces challenges, most notably the correlation between input signals and measurement noise caused by feedback~\cite{Vandenhof98CLissues, Forssell99Revisited}.

Classical methods for closed-loop identification are grouped into direct, joint input-output, and indirect approaches~\cite{Soderstrom88SysID}.
Direct methods estimate both the system and noise model independently of the feedback. They are consistent, i.e., can recover the true model with infinite data, if the noise model is accurate. However, real-world disturbances are often too complex to model precisely~\cite{Vandenhof98CLissues, Forssell99Revisited}.

Joint input-output methods first identify a system driven by white noise that generates the joint input-output signal, then simultaneously estimate the target system and controller~\cite{Chen24IOP}. 
While consistent even with imperfect noise models, these methods often suffer from theoretically unbounded variance~\cite{Abdalmoaty24SmallNoise} and high computational cost, especially for nonlinear controllers~\cite{Soderstrom88SysID}.
Indirect methods consider the closed-loop mapping and identify the system by exploiting knowledge of the controller \cite{Forssell99Revisited}. 
These methods parameterize the model to reformulate the problem, eliminating noise effects at the plant input and enabling the use of standard open-loop techniques~\cite{Vandenhof98CLissues, Forssell99Revisited}. This ensures consistency even when the noise model is misspecified, at least in the linear case.

One commonly used parameterization is the \emph{dual Youla} parameterization, which describes all systems stabilized by a given controller \cite{Hensen89DualYoula}. This formulation is the dual of the Youla-Kučera parameterization, which characterizes all controllers that stabilize a given system~\cite{Youla76Youla}. 
A key advantage of this approach is that it inherently ensures the identified model is stabilized by the controller. This aligns with the stability of the true system under the given closed-loop configuration~\cite{Forssell99Revisited}.
The dual Youla parameterization involves the coprime factorizations of both the controller and a nominal plant stabilized by it, along with Youla parameters that can be optimized to minimize the model's discrepancy from the target system.
For a linear plant and controller, the Youla parameters correspond to a stable rational transfer function, making optimization straightforward.

Given dual Youla's success in identifying linear systems, prior work has explored extending this parameterization to cases with nonlinear systems but linear controllers (see \cite{anderson1998youla} for a review).
However, two key challenges arise:
\begin{itemize}
    \item Coprime factorizations exist only for a subclass of nonlinear systems and are typically difficult to obtain even when they do exist~\cite{anderson1998youla}.
    \item The parameterization relies on stable operators, the nonlinear counterparts of the Youla parameter, and their inverses which are challenging to optimize over. A similar challenge has been noted in the context of learning stabilizing nonlinear controllers using the Youla parameterization \cite{outputSLS}.
\end{itemize}
As a result, existing methods do not lend themselves to a computationally tractable framework.

More recently, \cite{Maruta21ASF} proposed a tractable framework for closed-loop identification of nonlinear systems. While effective in their experiments, this method has two key limitations: 
\begin{enumerate*}[label=(\roman*)] 
    \item the closed-loop system of the learned model and controller may become unstable, and 
    \item it assumes a predefined parametric form for the system, limiting flexibility. 
\end{enumerate*}
This leaves closed-loop identification of expressive, nonlinear models a critical yet not sufficiently explored challenge.

\textbf{Contributions.}
We propose a closed-loop system identification framework for learning complex, potentially unstable models using neural networks while ensuring closed-loop stability with a given controller. Our approach is based on a parameterization inspired by the dual Youla framework, adapted to settings where both the system and controller may be nonlinear.
To this end, we first derive a complete characterization of all systems that can be stabilized by a given controller in terms of an operator in $\Lp$. Motivated by~\cite{outputSLS}, we then focus on a parametric subset of these operators, resulting in a computationally tractable optimization framework.
In the linear case, we establish the asymptotic consistency of our approach, aligning with existing results on the consistency of the dual Youla parameterization~\cite{Vandenhof98CLissues, Forssell99Revisited}.
Finally, numerical comparisons with direct baselines demonstrate the superiority of our method in capturing instability and achieving improved identification accuracy.

\subsection{Notation}
We represent sequences as matrices with infinitely many rows: $\trueStateMat = [\trueState{0}, \trueState{1}, \dots]^\top$, where $\trueState{t} \in \R^\dimState \ \forall t \in \N$.
The set of such sequences is $\ell^\dimState$, and $\trueStateMat \in \lp^\dimState \subset \ell^\dimState$ for $p \in \N$ if $\trueStateMat$ has bounded $p$-norm: 
    $\Vert \trueStateMat \Vert_p = \bigl( \sum_{t=0}^{\infty} \vert \trueState{t} \vert^p \bigr)^{1/p} < \infty$.
We denote truncation as $\trueState{j:i} = [\trueState{i}, \dots, \trueState{j}]^\top$ for $j \geq i$ and $\trueState{j:i} =\varnothing$ otherwise.

An operator $\upOp: \ell^\dimState \to \ell^\dimOut$ is a mapping between two sequences and is \textit{causal} if 
$\upOp(\trueStateMat) = \bigl(\upFunc{0}(\trueState{0}), \dots, \upFunc{t}(\trueState{t:0}), \dots\bigr)$, and \textit{strictly causal} if $\upOp(\trueStateMat) = \bigl(\upFunc{0}(\varnothing), \dots, \upFunc{t}(\trueState{t-1:0}), \dots\bigr)$, where $\upFunc{0}(\varnothing)$ is a vector in $\R^\dimOut$.
A causal operator $\upOp$ is:
\begin{itemize}
    \item $\Lp$-stable ($\upOp \in \Lp$) if $\upOp(\trueStateMat) \in \lp^\dimOut$ for all $\trueStateMat \in \lp^\dimState$,
    \item incrementally $\Lp$-stable if $\exists \, \gamma \in [0, \infty)$ such that $\forall \, \trueStateMat_1, \trueStateMat_2 \in \lp^\dimState$, 
        $\Vert \upOp(\trueStateMat_1) - \upOp(\trueStateMat_2)\Vert_p \leq \gamma \, \Vert\trueStateMat_1 - \trueStateMat_2\Vert_p$,
    where $\gamma$ is the incremental finite gain (i.f.g.) of $\upOp$.
\end{itemize}
The sets of causal and strictly causal $\Lp$-stable operators are denoted $\Lp^C$ and $\Lp^{SC}$, respectively.

\section{Plants stabilized by a given controller}
\subsection{Problem setup}
We consider a discrete-time, time-varying dynamical system that may be nonlinear and unstable, described by:
\begin{align} 
    \begin{cases}
        \trueNoisyOut{t} = \trueSysFunc{t} (\trueIn{t-1:0}) + \noise{t}, \quad t \in \N, 
        \\
        \trueNoisyOut{0} = \trueSysFunc{0} (\varnothing) + \noise{0},
    \end{cases}\label{eq:true_sys}
\end{align}
with input $\trueIn{t} \in \R^\dimIn$, nominal initial output $\trueSysFunc{0} (\varnothing) \in \R^\dimOut$, and output $\trueNoisyOut{t} \in \R^\dimOut$. The output is corrupted by independent noise $\noise{t}$. The system is strictly causal, meaning its output at time $t$ depends only on past inputs up to time $t-1$.

The model \eqref{eq:true_sys} is an input-output relationship, representing all strictly causal dynamical systems. The next example
shows how to derive it from a state-space model.
\begin{example}[From state-space to input-output models]
\label{ex:nonlin}
Consider a noise-free, strictly causal, time-invariant system in the state-space form:
\begin{align*}
    \begin{cases} 
        \trueState{t+1} = \dsFunc(\trueState{t}, \trueIn{t}) \,, \quad \trueState{0} = \bar{\trueState{}},\\
        \trueNoisyOut{t} = \outputFunc(\trueState{t}),
    \end{cases}
\end{align*}
where $\trueState{t} \in \R^\dimState$ is the state at time $t$ and $\bar{\trueState{}} \in \R^\dimState$ is the initial state.
By recursion, the functions $\trueSysFunc{t}$ in \eqref{eq:true_sys} are defined as: 
   \begin{align*}
        \trueSysFunc{0}(\varnothing) &= \outputFunc \bigl(\bar{\trueState{}}\bigr) , \; \trueSysFunc{1}(\trueIn{0}) = \outputFunc \bigl( \dsFunc (\bar{\trueState{}}, \trueIn{0})\bigr) , \; \dots\\
        \trueSysFunc{t}(\trueIn{t-1:0}) &= \outputFunc \bigl(\underbrace{ \dsFunc ( \dots ( \dsFunc }_{t \textit{ times}}(\bar{\trueState{}}, \trueIn{0}), \trueIn{1}), \dots, \trueIn{t-1})\bigr) \,, \; 
        \dots
    \end{align*}
These expressions show that $\trueSysFunc{t}$ depends explicitly on the initial state $\bar{\trueState{}}$ for all $t$, which is coherent with the definition of input-output models in the literature \cite{paice1996tac}. 
\end{example}
Moreover, linear operators with zero initial conditions represent
transfer functions, highlighting their universality in characterizing LTI systems~\cite{outputSLS}.

The system \eqref{eq:true_sys} can be expressed in operator form as:
\begin{align}
    \trueNoisyOutMat = \trueSys (\trueInMat) + \noiseMat, \label{eq:true_sys_op}
\end{align}
where $\trueSys$ is a strictly causal operator such that $\trueSys (\trueInMat) = [\trueSysFunc{0} (\varnothing), \trueSysFunc{1} (\trueIn{0}), \dots]^\top$, $\trueInMat = [\trueIn{0}, \trueIn{1}, \dots]^\top$, etc.
\begin{figure}[t]
    \centering
    \includegraphics[width=.9\linewidth]{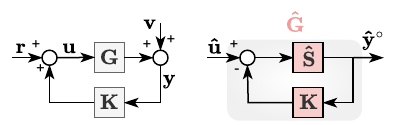}
    \caption{\textit{Left:} The closed-loop of system $\trueSys$ and controller $\cont$, with  input $\trueInMat$, noisy output $\trueNoisyOutMat$, excitation signal $\excMat$, and noise $\noiseMat$.
    \textit{Right:} In the ICI framework, the model $\approxSys$ is an interconnection of a copy of the controller $\cont$ and a trainable operator $\approxOp \in \Lp^{SC}$. The model input and output are denoted by $\approxInMat$ and $\approxOutMat$, respectively.}
    \label{fig:true_CL_Ghat}
\vspace{-3mm}
\end{figure}
We assume the system is controlled by a causal dynamic output feedback controller $\cont$. In addition to the feedback loop, a persistently exciting (PE) signal $\exc{t} \in \R^\dimIn$ is applied to the system.
The resulting closed-loop system, illustrated in the left panel of \autoref{fig:true_CL_Ghat}, is:
\begin{align} 
    \begin{cases} 
        \trueNoisyOutMat = \trueSys (\trueInMat) + \noiseMat,
        \\ 
        \trueInMat = \excMat + \cont(\trueNoisyOutMat),  
    \end{cases} \label{eq:true_cl}
\end{align}
where $\cont (\trueNoisyOutMat) = [\contFunc{0} (\trueNoisyOut{0}), \contFunc{1} (\trueNoisyOut{1:0}), \dots]^\top$ and $\excMat = [\exc{0}, \exc{1}, \dots]^\top$.
We assume the closed-loop system formed by $\trueSys$ and $\cont$ is $\Lp$-stable, as defined below.
\begin{definition}[Closed-loop $\Lp$-stability \cite{PerfBoost}]\label{def:stability}
    The closed-loop of $\trueSys$ and $\cont$, as shown in the left panel of \autoref{fig:true_CL_Ghat}, is $\Lp$-stable if for every $\excMat \in \lp$ and $\noiseMat \in \lp$, it holds that $\trueInMat \in \lp$ and $\trueNoisyOutMat \in \lp$.
    Equivalently, the mapping $(\excMat, \noiseMat) \mapsto (\trueInMat, \trueNoisyOutMat)$ is an $\Lp$ operator.
\end{definition}

Our goal is to learn a dynamical system $\approxSys$ that approximates $\trueSys$, given knowledge of the controller $\cont$. Crucially, we allow $\approxSys$ to be unstable, enabling the identification of an unstable $\trueSys$, while ensuring that the closed-loop system formed by $\approxSys$ and $\cont$ remains stable. This requirement maintains consistency with the fact that $\cont$ stabilizes $\trueSys$.

\subsection{Internal controller identification framework}\label{subsec:ici}
Learning over the space of systems $\approxSys$ stabilized by controller $\cont$ is challenging, as this space is generally nonconvex. 
To address this, we parameterize all and only systems stabilized by $\cont$ using a dual formulation of the Internal Model Control (IMC) framework \cite{IMC}, which we call Internal Controller Identification (ICI). Here, $\approxSys$ is structured as an interconnection of controller $\cont$ and a trainable strictly causal operator $\approxOp \in \Lp^{SC}$. This transforms the problem of learning $\approxSys$ into optimizing $\approxOp$ within $\Lp^{SC}$ operators, enabling unconstrained optimization \cite{outputSLS, PerfBoost}.
The corresponding model is illustrated in the right panel of \autoref{fig:true_CL_Ghat}, where the noise-free output $\approxOut{t} \in \R^\dimOut$ for an input $\approxIn{t} \in \R^\dimIn$ is given by:
\begin{equation} 
    \approxOutMat = \approxOp \bigl( \approxInMat - \cont (\approxOutMat) \bigr). \label{eq:ici}
\end{equation}
\begin{remark}[Recursive implementation of $\approxOp$]\label{remark:recursive}
    Since $\approxOp$ is strictly causal, the output $\approxOut{t}$ depends on past outputs only up to time $t-1$, i.e., 
    $
        \approxOut{t} = \approxOpFunc{t} \bigl( \approxIn{t-1:0} - \contFunc{t-1:0} 
        (\approxOut{t-1:0}) \bigr).
    $
    This structure enables a recursive computation of $\approxOut{t}$, eliminating the need to solve the implicit nonlinear equation~\eqref{eq:ici} at each time step.
    Additionally, if $\approxOp$ and $\cont$ are input-output operators associated with dynamical systems (see \autoref{ex:nonlin}), their state-space representations can be used to implement \eqref{eq:ici} as a Markovian dynamical system, thereby avoiding the need to store the full history of $\approxIn{t-1:0}$ and $\approxOut{t-1:0}$.
\end{remark}

The following theorem establishes that a given stable controller, $\cont\in\Lp$, can stabilize a system if and only if the system can be represented within the ICI framework.
\begin{theorem}\label{theo:ici}
    Let the controller $\cont \in \Lp^\textit{C}$ be an incrementally stable operator. Under the ICI framework, as shown in the right panel of \autoref{fig:true_CL_Ghat} and described by \eqref{eq:ici}, the following holds: 
    \begin{enumerate}[label=\textcolor{darkgray}{\textbf{T.\arabic*}}]
        \item \label{item:T1} Let $\approxOp:\lp^\dimIn \rightarrow \lp^\dimOut$ be an $\Lp^\textit{SC}$ operator and consider the ICI model $\approxSys$ shown in the right plot of \autoref{fig:true_CL_Ghat}. Then, $\approxSys$ is stabilized by $\cont$ in the sense of \autoref{def:stability}.
        \item \label{item:T2} Let $\trueSys$ in \autoref{fig:true_CL_Ghat} be a strictly causal system such that the closed-loop system is stable according to \autoref{def:stability}. Then, there exists  $\approxOp \in \Lp^\textit{SC}$ such that $\trueSys=\approxSys$, where $\approxSys$ is constructed as in the right panel of \autoref{fig:true_CL_Ghat}.
    \end{enumerate}
\end{theorem}
The proof can be found in the appendix of~\cite{Boroujeni2025Neural}.

Since $\trueSys$ is stabilized by $\cont$, by point \ref{item:T2} of \autoref{theo:ici}, it admits an ICI representation in terms of an operator $\trueOp$, which we refer to as the \emph{true ICI} parameterization.
The true operator $\trueOp$ serves as a benchmark for evaluating algorithms that learn $\approxOp$. Specifically, we assess a learning algorithm's effectiveness by its ability to recover $\trueOp$ in \autoref{sec:analysis}.

\section{Training procedure}\label{sec:training}
In this section, we discuss learning $\approxSys$ from data to approximate the unknown system $\trueSys$. The dataset is generated by applying $\numSamples \in \mathbb{N}$ excitation signals, each over a horizon $\horizon \in \mathbb{N}$, to the closed-loop system in the left plot of \autoref{fig:true_CL_Ghat}, and recording the resulting system inputs and noisy outputs, defined as:
\begin{equation*}
    \dataset 
    = 
    \bigl\{ ( \exc{\horizon-1:0}^\counter, \trueIn{\horizon-1:0}^\counter, \trueNoisyOut{\horizon-1:0}^\counter ) \bigr\}_{\counter=0}^{\numSamples-1}
    ,
\end{equation*}
where $\exc{\horizon-1:0}^\counter$ represents the first $\horizon$ time steps of $\exc{}$ in the experiment $\counter \in \{0, \dots, \numSamples-1\}$ (similarly for $\trueIn{}$ and $\trueNoisyOut{}$).

Using this dataset, we explore three identification strategies: two direct and one indirect. These strategies aim to minimize the discrepancy between the measured and predicted outputs, $\trueNoisyOutData$ and $\approxOutData$, based on the mean square error (MSE) cost function:
\begin{align}
    \cost \bigl(\dataset, \approxOutData \bigr) = \frac{1}{\numSamples} \sum_{\counter=0}^{\numSamples-1} \frac{1}{\horizon} \sum_{t=0}^{\horizon-1} \Vert \trueNoisyOut{t}^\counter - \approxOut{t}{^\counter} \Vert_2^2. \label{eq:cost}
\end{align}
The key difference between the learning strategies lies in how the predicted output is obtained, as outlined below.

\subsection{Direct identification}
In direct identification, the measured inputs $\trueInData$ are applied to the model $\approxSys$, without considering the controller $\cont$. This effectively treats the data as if collected in an open-loop setting. We use this approach to learn both a general $\approxSys$ (not necessarily in the ICI formulation) and $\approxOp$ within the ICI framework, leading to the following strategies.

\paragraph*{\textbf{\strat{1}}} \label{strat1} 
Direct identification of $\approxSys$, formulated as:
\begin{align*} 
    \approxSys^{\textit{\dirIDSup}} = \argmin_{\approxSys}\;& 
    \cost \bigl(\dataset, \approxOutData \bigr),
    \\ 
    \textit{s.t.} \;& 
    \approxOutMat{^\counter} = \approxSys (\trueInMat^\counter) \quad  \forall \counter
    . 
\end{align*}
The resulting model $\approxSys^{\textit{dir-ID}}$ is not necessarily stabilized by $\cont$, neglecting the knowledge that the true system $\trueSys$ is stabilized by $\cont$.  
Nonetheless, we use this strategy as a baseline due to its widespread use. To ensure closed-loop stability, the next two strategies leverage the ICI framework.

\paragraph*{\textbf{\strat{2}}} \label{strat2} 
Direct identification of $\approxOp$ in the ICI framework, leading to:

\resizebox{.98\linewidth}{!}{
  \begin{minipage}{\linewidth}
    \begin{align*}
        \approxOp^{\textit{\dirICISup}} = \argmin_{\approxOp \in \Lp^\textit{SC}}\;& \cost \bigl(\dataset, \approxOutData \bigr),
        \\
        \textit{s.t.} \;& 
        \approxOutMat{^\counter} = \approxOp \bigl( \trueInMat^\counter - \cont(\approxOutMat{^\counter}) \bigr) \quad  \forall \counter, 
    \end{align*}
    \end{minipage} }
where the constraint follows from substituting $\approxInMat = \trueInMat$ in \eqref{eq:ici}. Recall that $\approxOutMat$ can be computed recursively by \autoref{remark:recursive}. 
The corresponding model, $\approxSys^{\textit{\dirICISup}}$, is then defined as the closed-loop system of $\approxOp^{\textit{\dirICISup}}$ and $\cont$. Notably, $\approxSys^{\textit{\dirICISup}}$ is stabilized by $\cont$, as guaranteed by \autoref{theo:ici}.

While direct methods lead to tractable optimization problems, they have two fundamental drawbacks.
\begin{enumerate}[label=\textcolor{darkgray}{\textbf{D.\arabic*}}]
    \item \label{item:D1}Direct methods apply only when $\approxSys$ is stable~\cite{Forssell99Revisited} since the cost function $\cost$ is evaluated using open-loop predictions of $\approxSys$, which may diverge if the model is unstable.
    \item \label{item:D2} Direct methods can yield inconsistent estimates, meaning that the true $\trueSys$ is not recovered even with infinite data \cite{Soderstrom88SysID, Vandenhof98CLissues, Goodwin2002Bias}. This inconsistency arises because the noise $\noiseMat$ affects both the trainable operator’s input, $\trueInMat$ in \nameref{strat1} and $\trueInMat - \cont (\approxOutMat)$ in \nameref{strat2}, and the measured output $\trueNoisyOutMat$, due to the feedback loop present during data collection, as illustrated in the left plot of \autoref{fig:true_CL_Ghat}. We further discuss this issue in \autoref{sec:analysis}.
\end{enumerate}

\subsection{Indirect identification}
\begin{figure}
    \centering
    \hspace{-2pt}\includegraphics[width=0.7\linewidth]{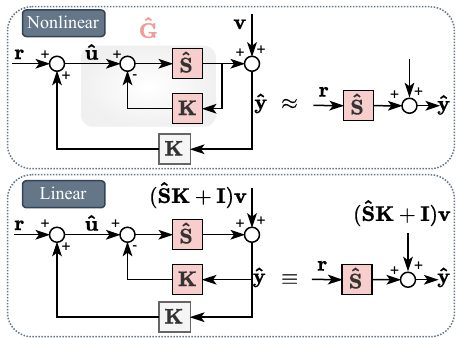}
    \caption{\textit{Top:} Closed-loop system of $\approxSys$ and $\cont$ (left) and its linearized approximation (right).
    \textit{Bottom:} In the linear case, noise injection can be shifted to enable loop cancellation, yielding the exact simplified form on the right, consistent with \eqref{eq:linearized_out}.}
    \label{fig:approx-cl-all}
    \vspace{-3mm}
\end{figure}

Indirect methods first identify the closed-loop system of $\approxSys$ and $\cont$, using the excitation signal $\excMat$ as input, as shown in the top-left plot of \autoref{fig:approx-cl-all}. They then leverage the known controller $\cont$ to extract a system model from the identified closed-loop system \cite{Soderstrom88SysID}.
Since learning is based on the closed-loop system of $\approxSys$ and $\cont$, these methods avoid direct open-loop simulation of $\approxSys$, thus addressing the problem \ref{item:D1}. Additionally, indirect methods assume a particular model parameterization in order to remove the noise contribution on the input of the trainable model, resolving issue \ref{item:D2} \cite{Vandenhof98CLissues}. 

In the following, we develop an indirect strategy leveraging the ICI framework. The proposed scheme is inspired by indirect methods based on the dual Youla parameterization (see, e.g., \cite{anderson1998youla}). 
As discussed in \autoref{sec:intro}, classical Youla-based methods are restricted to linear controllers, require intractable coprime factorizations for nonlinear systems, and involve operators that hinder computational optimization. By utilizing the ICI framework and recent techniques for parameterizing stable operators, we circumvent coprime factorizations and establish a computationally tractable framework. 

We begin by computing the noisy output $\approxNoisyOutMat$ resulting from applying $\excMat$ to the closed-loop system of $\approxSys$ and $\cont$, as illustrated in the top-left plot of \autoref{fig:approx-cl-all}:
\begin{align}
    \approxNoisyOutMat &= \approxOp \bigl(\approxInMat - \cont(\approxNoisyOutMat - \approxNoiseMat) \bigr) + \approxNoiseMat
    \nonumber \\
    &= \approxOp \bigl(\excMat + \cont(\approxNoisyOutMat) - \cont(\approxNoisyOutMat - \approxNoiseMat) \bigr) + \approxNoiseMat. \label{eq:strat3_y}
\end{align} 
Next, following \cite{anderson1998youla}, we move the noise outside the operand of $\approxOp$ in \eqref{eq:strat3_y} by linearizing $\approxOp$ and $\cont$ around $\excMat$ and $\approxNoisyOutMat$, respectively. This results in:
\begin{align}
    \approxNoisyOutMat 
    &\approx 
    \approxOp \bigl( \excMat \bigr) + \approxOp_L \bigl( \cont(\approxNoisyOutMat) - \cont(\approxNoisyOutMat - \approxNoiseMat) \bigr) + \approxNoiseMat
    \nonumber\\
    &\approx 
    \approxOp (\hspace{-24pt}\underbrace{\excMat}_{\textit{noise-independent signal}} \hspace{-24pt})
    + 
    \underbrace{\approxOp_L \bigl( \cont_L (\approxNoiseMat) \bigr) + \approxNoiseMat}_{\textit{noise}},
    \label{eq:linearized_out}
\end{align}
where $\approxOp_L$ and $\cont_L$ denote the linearized versions of $\approxOp$ and $\cont$, around $\excMat$ and $\approxNoisyOutMat$, respectively. If $\approxOp$ and $\cont$ are linear, the approximation holds exactly, with $\approxOp_L = \approxOp$ and $\cont_L = \cont$. 

\begin{remark}[Validity of linear approximations]
    The above linearization of $\cont$ holds well at high output signal-to-noise ratio (SNR), where $\approxNoisyOutMat$ has significantly higher power than $\noiseMat$.\footnote{The power of a signal $\mathbf{y}$ is defined as $\lim_{T \to \infty} \frac{1}{T} \sum_{t=0}^T \Vert y_t \Vert_2^2$ \cite{oppenheim_signals_1983}.} Similarly, $\approxOp$ is well-approximated when $\excMat$ has a much higher power than $\cont(\approxNoisyOutMat) - \cont(\approxNoisyOutMat - \approxNoiseMat)$. Since $\cont$ is incrementally stable, this condition simplifies to a high excitation SNR, where the power of $\excMat$ is much greater than that of $\noiseMat$.
\end{remark}

According to \eqref{eq:linearized_out}, the closed-loop response of the system formed by $\approxSys$ and $\cont$ to $\excMat$ can be approximated by passing $\excMat$ directly through $\approxOp$. Intuitively, the internal and external feedback loops nearly cancel out, allowing $\excMat$ to propagate primarily through $\approxOp$ (see the top-right plot of \autoref{fig:approx-cl-all}). 
This approximation is exact in two cases. First, in the absence of noise, as is immediately evident from the top-left plot of \autoref{fig:approx-cl-all}. Second, when $\approxOp$ and $\cont$ are linear, rearranging the top-left plot in \autoref{fig:approx-cl-all} by shifting the noise yields the bottom-left plot. As shown, the feedback loops cancel out, leading to the simplified diagram on the right, which aligns with \eqref{eq:linearized_out}.

In light of \eqref{eq:linearized_out}, we introduce the third strategy.
\paragraph*{\textbf{\strat{3}}} \label{strat3} Indirect identification of $\approxOp$ in the ICI framework via output linearization, formulated as:
\begin{align*}
    \approxOp^{\textit{\indirICISup}} = \argmin_{\approxOp \in \Lp^\textit{SC}}\;& 
    \cost \bigl(\dataset, \approxOutData \bigr),
    \\ 
    \textit{s.t.} \;& 
    \approxOutMat{^\counter} = \approxOp (\excMat^\counter) \quad \forall \counter.
\end{align*}
The corresponding model $\approxSys^{\textit{\indirICISup}}$ is then defined as the closed-loop system formed by $\approxOp^{\textit{\indirICISup}}$ and $\cont$.

\begin{remark}[Open-loop equivalence]\label{remark:openloop}
    We observe from \eqref{eq:linearized_out} that (i) $\approxNoisyOutMat$ can be approximated as $\approxOp$ applied to $\excMat$, which is independent of the noise, plus an additive noise term, and (ii) $\approxOp$ is a stable operator. As a result, \nameref{strat3} is a \emph{standard open-loop} problem, eliminating the instability (\ref{item:D1}) and correlation (\ref{item:D2}) issues of direct methods. This equivalence is well-known for classic indirect approaches, such as the dual Youla parameterization~\cite{Forssell99Revisited}.
\end{remark}

Following \autoref{remark:openloop}, we can use any open-loop method, such as subspace identification or instrumental variables (see~\cite{Soderstrom88SysID} for an overview), to obtain the system model. Here, we minimize the least squares cost in \eqref{eq:cost}, aligning with the prediction error method~\cite{Soderstrom88SysID}. Moreover, \autoref{remark:openloop} allows us to directly leverage the consistency properties of open-loop identification, as discussed in \autoref{sec:analysis}.

\subsection{Parameterizing the operator \texorpdfstring{$\approxOp$}{Lg}}
To implement the ICI framework, we must parameterize the operator $\approxOp \in \Lp^\textit{SC}$. This can be achieved using various families of parameterized $\Lp^\textit{SC}$ operators, such as Recurrent Equilibrium Networks (RENs)~\cite{revay2023recurrent}, certain classes of State-Space Models (SSMs)~\cite{gu2022efficiently}, and neural network representations of Hamiltonian systems~\cite{zakwan2024neural}. These operators all admit a state-space form, enabling recursive computation of the output of $\approxSys$ (see \autoref{remark:recursive}). In our experiments, we follow~\cite{PerfBoost, Boroujeni24PAC} and employ RENs, which define a broad class of $\Ltwo$ operators and embed arbitrarily deep neural networks~\cite{revay2023recurrent}. 

For any chosen operator family, we denote the parameterized operator and the associated model by $\approxOp^{\params}$ and $\approxSys^{\params}$, with learnable parameters $\params \in \R^\dimParams$. 
Importantly, these families are constructed so that $\approxOp^{\params} \in \Lp^\textit{SC}$ for all $\params \in \R^\dimParams$.
Consequently, optimizing over $\approxOp \in \Lp^\textit{SC}$ in \nameref{strat2} and \nameref{strat3} reduces to an unconstrained optimization over $\params \in \R^\dimParams$, which can be efficiently solved using gradient descent methods.

\section{Statistical analysis in the linear case}\label{sec:analysis}

The identification procedure is inherently stochastic, as it depends on the realizations of noise and excitation signals in the dataset. A central question is whether the learned models $\approxSys$ and $\approxOp$ asymptotically converge in probability to the true system $\trueSys$ or the true $\trueOp$, defined in \autoref{subsec:ici}, as the sample size grows. This notion of convergence is formalized through the concept of consistency:\footnote{In some classic system identification literature, the term unbiasedness is also used to describe consistency (e.g., \cite{Forssell99Revisited}). To align with machine learning terminology, we use consistency throughout.}
\begin{definition}[Consistency \cite{Vandenhof98CLissues}]
    An estimate $\approxSys$ (or $\approxOp$) obtained based on a dataset of $\numSamples$ sequences, each of horizon $\horizon$, is consistent if it converges to the true value $\trueSys$ (or $\trueOp$) with probability $1$ as the sample size grows indefinitely, i.e.,
\begin{align*}
    \lim_{\numSamples\, \horizon \to \infty} \approxSys = \trueSys,
    \quad \text{or} \quad
    \lim_{\numSamples\, \horizon \to \infty} \approxOp = \trueOp.
\end{align*}
\end{definition}
Since $\approxOp$ and $\approxSys$ can be uniquely determined from one another, the consistency of $\approxOp$ is equivalent to that of $\approxSys$.

We discuss the consistency of the proposed strategies when $\cont$, $\trueSys$, and $\approxSys$ are linear. 
Throughout this section, we assume that the excitation signal is quasi-stationary and sufficiently rich to identify the plant dynamics \cite{ljung1999system, Forssell99Revisited}.

The consistency of direct methods (\nameref{strat1} and \nameref{strat2}) can only be analyzed for a stable $\approxSys$ (see \ref{item:D1}). Even in this case, it depends heavily on the noise model. Let the noise $\noiseMat$ in \eqref{eq:true_sys_op} be expressed as $\noiseMat = \noiseModel (\whiteNoiseMat)$, where $\whiteNoiseMat$ is white noise and $\noiseModel$ is a linear noise model.
Direct methods typically estimate both the target system $\trueSys$ and the operator $\noiseModel$. It is well-established that a consistent estimate of $\trueSys$ can only be obtained if $\noiseModel$ is either known or exactly learned \cite{Vandenhof98CLissues, Forssell99Revisited, Maruta21ASF, Ljung99Alternative}. However, many real-world disturbances exhibit colored noise characteristics \cite{Bak87PinkNoise}, which cannot be captured by a finite-order noise model $\noiseModel$. As a result, direct identification methods are often inconsistent in practice \cite{Ljung99Alternative, Maruta21ASF}.

Next, we discuss the consistency of \nameref{strat3}. As noted in \autoref{remark:openloop}, \nameref{strat3} is equivalent to identifying $\approxOp$ in an open-loop setting.
Consequently, consistency results from open-loop identification apply directly, as stated below.
\begin{proposition}[Corollary 6 in \cite{Forssell99Revisited}]\label{prop:consistency}
    Consider the stable closed-loop system given by \eqref{eq:true_cl}, where the system $\trueSys$ and the controller $\cont$ are linear.
    Suppose the dataset $\dataset$ consists of $\numSamples$ sequences, each of horizon $\horizon$, collected under quasi-stationary persistent excitation $\excMat$.
    Let $\approxOp^{\params}$ be the estimate obtained by \nameref{strat3} using $\dataset$, optimized over a parametric family of $\Lp^\textit{SC}$ operators. Then, both $\approxOp^{\params}$ and the corresponding $\approxSys^{\params}$ are consistent, provided that the true $\trueOp$, as introduced in \autoref{subsec:ici}, belongs to the chosen family.
\end{proposition}

The consistency of \nameref{strat3} only requires the noise to be independent of the excitation signal, without imposing any additional assumptions on the noise.

\autoref{prop:consistency} requires that the true $\trueOp$ belongs to the chosen parametric set. If this set covered the entire $\Lp^\textit{SC}$ space, consistency would be guaranteed. However, since most parametric families only span a subset of it, $\trueOp$ may fall outside, leading to inconsistency. This is a common trait in the analysis of learning algorithm consistency \cite{Forssell99Revisited}, which can be addressed by using a more flexible parameterization or increasing the dimensionality of the parameters, $\dimParams$.

\section{Experiments}\label{sec:experiments}
We consider two nonlinear systems: \textit{i)} an unstable scalar system stabilized with nonlinear feedback, and \textit{ii)} a stable planar robot, controlled to navigate to a target location. We compare our framework (\nameref{strat3}) against the baselines (\nameref{strat1} and \nameref{strat2}), showing the effectiveness of our approach in both stable and unstable settings. 
\footnote{The code is available on \url{https://github.com/DecodEPFL/Nonlinear_system_identification}.} 

\textbf{Training datasets.} For both experiments, the training dataset $\dataset$ consists of $\numSamples = 40$ trajectories, each starting from a fixed initial condition, with a horizon of $\horizon = 100$. The system input is given by $\trueIn{t} = \contFunc{t}(\trueNoisyOut{t:0}) + \exc{t}$, where the controller $\contFunc{t}$ will be specified for each experiment. The excitation signal $\exc{t}$ is independently sampled at each time step from a zero-mean Gaussian distribution, $\exc{t} \sim \mathcal{N}(0, \sigma^2 \eye)$, ensuring sufficient exploration of the system dynamics \cite{ljung1999system}. The value of $\sigma$ varies across experiments.

\textbf{Evaluation criteria.} We assess all strategies on a test set of $100$ trajectories of length $\horizon=100$, using the same excitation and noise distributions as in training. Four criteria are used: \emph{open-loop (OL) MSE}, \emph{OL coefficient of determination (R\textsuperscript{2})}, \emph{closed-loop (CL) MSE}, and \emph{CL R\textsuperscript{2}}.
In the open-loop setup, the excitation signal is applied to both the true system and the learned model, and the MSE between their outputs is computed. 
Since MSE is scale-dependent, we also report the coefficient of determination (R\textsuperscript{2}), which 
provides a standardized measure of how well the model explains the variance in the dependent variable~\cite{james2013introduction}:
$
    R^2 = 1 - \frac{\sum_{\counter=1}^\numSamples  \sum_{t=1}^\horizon 
        \Vert \trueNoisyOut{t}^\counter - \approxOut{t}{^\counter} \Vert_2^2
    }{\sum_{\counter=1}^\numSamples \sum_{t=1}^\horizon 
        \Vert \trueNoisyOut{t}^\counter - \bar{\trueNoisyOut{}} \Vert_2^2
    }$.
It ranges from $-\infty$ to $1$ (values closer to $1$ indicate a better fit) and $\bar{\trueNoisyOut{}}$ is the overall mean of $\trueNoisyOut{}$.

While OL criteria directly measure model accuracy, they do not apply to unstable systems, as their open-loop outputs diverge. 
Therefore, we also assess closed-loop performance. The same excitation signals and noise sequences are applied to both the true closed-loop system (comprising $\trueSys$ and controller $\cont$) and the learned closed-loop system (formed by $\approxSys$ and $\cont$), and the CL MSE and R\textsuperscript{2} are computed. Since closed-loop outputs remain bounded when the controller stabilizes the system, CL criteria allow meaningful comparisons even for unstable plants. 
In the sequel, we repeat all experiments with $50$ different random seeds and report $95\%$ confidence intervals for each evaluation criterion. 

\textbf{Model details.}
We instantiate $\approxSys$ in \nameref{strat1} and $\approxOp$ in \nameref{strat2} and \nameref{strat3} as RENs \cite{revay2023recurrent}. In all cases, the REN architecture comprises a state vector in $\R^8$ and $8$ hidden layers, resulting in a total of $744$ trainable parameters. This configuration provides sufficient modeling flexibility.

\subsection{Scalar unstable system}
We consider the following state-space scalar system:
\begin{align}
    \begin{cases} 
        \trueState{t+1} = \trueState{t}^2 + 1 + \trueIn{t}, \quad \trueState{0} = \bar{\trueState{}},\\
        \trueNoisyOut{t} = \trueState{t} + \noise{t},
    \end{cases} \label{eq:exp_toy_sys}
\end{align}
where $\trueState{t}$ is the state and the excitation signal standard deviation is $\sigma = 0.5$. The noise $\noise{t}$ is sampled from a truncated normal distribution with a zero mean, standard deviation of $0.1$, and bounded in the interval $(-0.25,0.25)$. 

The system in \eqref{eq:exp_toy_sys} is unstable, making closed-loop identification necessary. We use the controller
$
    \contFunc{t}(\trueNoisyOut{t:0}) = -\trueNoisyOut{t}^2-1+0.5\trueNoisyOut{t}
$, resulting in the closed-loop system:
\begin{align*}
    \begin{cases} 
        \trueState{t+1} = (0.5 - 2 \noise{t}) \trueState{t} + 0.5 \noise{t} - \noise{t}^2 + \exc{t}, \quad \trueState{0} = \bar{\trueState{}},\\
        \trueNoisyOut{t} = \trueState{t} + \noise{t}.
    \end{cases}
\end{align*}
Since $\noise{t} \in (-0.25, 0.25)$, the output converges on average for any initial condition and does not grow infinitely.

\begin{figure}
    \centering
    \begin{subfigure}[b]{0.48\linewidth}
        \centering
        \includegraphics[width=\linewidth]{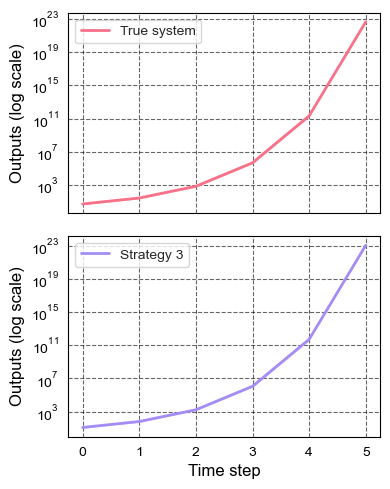}
    \end{subfigure}
    \hfill
    \begin{subfigure}[b]{0.48\linewidth}
        \centering
        \includegraphics[width=\linewidth]{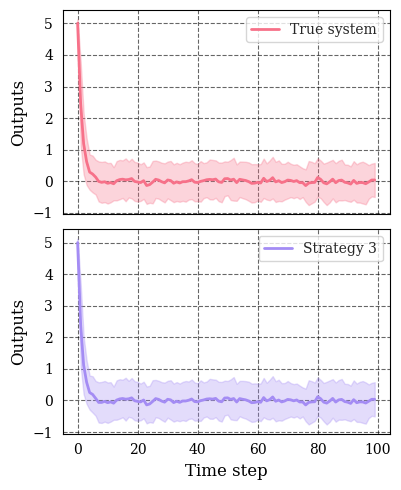}
    \end{subfigure}
    \caption{
    Comparison of system and model outputs in the scalar unstable experiment. \textit{Top:} true system output; \textit{Bottom:} output of the model learned by \nameref{strat3}; \textit{Left:} open-loop; \textit{Right:} closed-loop. Solid lines show the average output across multiple realizations of noise and excitation signals and shaded regions indicate $95\%$ confidence bounds. The open-loop confidence bounds are not visible due to their much smaller scale relative to the output. The learned model mirrors the true system’s stability properties: it is unstable in open-loop but stable in closed-loop. Closed-loop responses closely match, with a low MSE of $0.0034 \pm 0.0006$.
    }
    \label{fig:comparison}
    \vspace{-3mm}
\end{figure}

\textbf{Results.}
Since the system is unstable, an unstable model $\approxSys$ must be used, making the direct strategies (\nameref{strat1} and \nameref{strat2}) inapplicable (see \ref{item:D1}). For \nameref{strat3}, we compare the OL and CL responses against the true system in \autoref{fig:comparison}.
In all plots, solid lines represent the average output over noise and excitation realizations, while shaded regions indicate the confidence bounds.
The OL plots highlight \nameref{strat3}’s ability to capture unstable dynamics, as the model’s outputs diverge significantly within just $5$ time steps. The OL confidence bounds are not visible due to the large output scale. As a result of these extreme values, OL criteria are not computable. The CL plots show a high similarity between the CL trajectories of the true and identified models, with a low CL MSE of $0.0034 \pm 0.0006$. We also observe that the output converges to zero.
This experiment showcases the effectiveness of our proposed approach in learning an unstable system.

\subsection{Point-mass robot}
We follow the experimental setup of \cite{PerfBoost}, modeling a planar robot as a point mass $m \in \R_+$. The state is the robot's position $\trueState{1,t} \in \R^2$ and velocity $\trueState{2,t} \in \R^2$, while the output is a noisy position measurement. The input $\trueIn{t}$ is a planar force driving the motion according to the state-space equations:
\begin{align*}
\begin{cases}
    \begin{bmatrix}
        \trueState{1,t+1} \\ \trueState{2,t+1}
    \end{bmatrix}
    = 
    \begin{bmatrix}
        \trueState{1,t} \\ \trueState{2,t}
    \end{bmatrix}
    + T_s
    \begin{bmatrix}
        \trueState{2,t} \\
        \frac{1}{m} \bigl(-C(\trueState{2,t}) + \trueIn{t} \bigr)
    \end{bmatrix}, \\
    \trueNoisyOut{t} = \trueState{1,t} + \noise{t}, 
\end{cases}
\end{align*}
where $T_s > 0$ is the sampling time and $\noise{t}\sim \mathcal{N}(0, 0.1)$. 
Moreover, the function $C: \R^2 \to \R^2$ models nonlinear drag forces, such as air or water resistance.
We set $C(\trueState2) = b_1 \trueState2 - b_2  \Vert \trueState2 \Vert_2 \, \trueState2$ for some $0 < b_2 < b_1$.

This system is stable due to the presence of drag forces. However, we assume the robot is controlled by a proportional controller that drives it toward a predefined target position $\trueNoisyOut{}^* = \zeroMat \in \R^2$, potentially for economic reasons. The control law is given by
$\contFunc{t}(\trueNoisyOut{t:0}) = \operatorname{diag}(\kappa_1, \kappa_2) (\trueNoisyOut{}^* - \trueNoisyOut{t})$, with $\kappa_1, \kappa_2 \in \R_+$. Specific parameter values for the system and the controller ($m$, $b_1$, $b_2$, $\kappa_1$, $\kappa_2$) are provided in the codebase.

\begin{figure}
    \includegraphics[width=\linewidth]{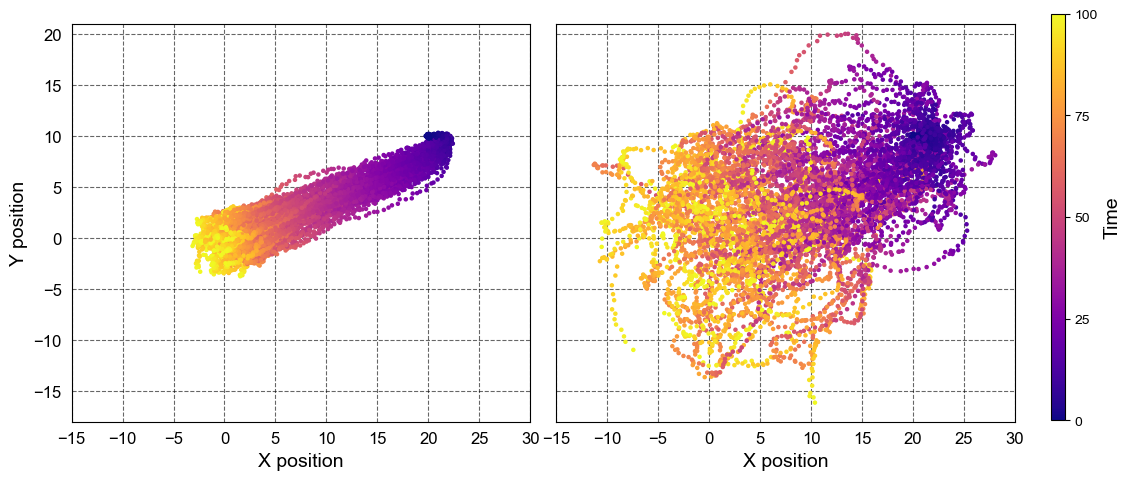}
    \caption{Training trajectories in the robot experiment. \textit{Left:} excitation standard deviation $\sigma = 10$; \textit{Right:} $\sigma = 50$. The robot has an initial velocity of $[10, 0]^\top$ and operates in a loop with a controller that drives it to the origin. As expected, the robot deviates less from its intended trajectory under lower excitation, and more significantly when the excitation is higher. Colors indicate time progress.}
    \label{fig:robots_dataset}
    \vspace{-3mm}
\end{figure}

We analyze the impact of the excitation signal's standard deviation, $\sigma$. First, we consider a small $\sigma = 10$. The training trajectories are shown in \autoref{fig:robots_dataset}, where the robot starts with an initial velocity of $[10, 0]^\top$. With low excitation, the trajectories remain smooth, allowing the robot to reach its target with minimal deviation. Next, we apply a high $\sigma = 50$. While larger excitation improves identification (as shown in the results section), \autoref{fig:robots_dataset} illustrates that it also introduces undesired deviations during the robot’s movement.

\begin{table*}
\caption{Evaluation of different strategies on the point-mass robot experiment, considering two excitation signal levels: $\sigma = 10$ and $\sigma = 50$. Each experiment is repeated $50$ times with different random seeds, and results are shown with $95\%$ confidence intervals. The proposed \nameref{strat3} consistently outperforms baselines across all settings and metrics. Colored cells indicate the best results.}
\label{tab:robot}
\begin{center}
\resizebox{0.7\textwidth}{!}{
\begin{tabular}{c|ccccc}
\toprule
& \textbf{Strategy} & {OL MSE} & {CL MSE} & {OL R\textsuperscript{2}} & {CL R\textsuperscript{2}} 
\\
\midrule
\parbox[t]{1.5mm}{\multirow{3}{*}{\rotatebox[origin=c]{90}{$\sigma=10$}}}&
\nameref{strat1} & $17.6847 \pm 0.0992$ & $0.4800 \pm 0.0090$ & $0.8368 \pm 0.0007$ &  $0.9902 \pm 0.0002$
\\
& \nameref{strat2} & $11.5468 \pm 0.0674$ & $0.3916 \pm 0.0086$ & $0.8934 \pm 0.0005$ &  $0.9920 \pm 0.0002$
\\
& \nameref{strat3} & \cellcolor{\bestres}$6.7351 \pm 0.0586$ & \cellcolor{\bestres}$0.2398 \pm 0.0049$ & \cellcolor{\bestres}$0.9378 \pm 0.0004$ &  \cellcolor{\bestres}$0.9951 \pm 0.0001$
\\
\midrule
\parbox[t]{2.5mm}{\multirow{3}{*}{\rotatebox[origin=c]{90}{$\sigma=50$}}}&
\nameref{strat1} & $16.2409 \pm 0.4997$ & $6.2671 \pm 0.1598$ & $0.8842 \pm 0.0029$ & $0.9122 \pm 0.0019$
\\
& \nameref{strat2} & $3.8891 \pm 0.1109$ & $1.8150 \pm 0.0258$ & $0.9723 \pm 0.0006$ & $0.9746 \pm 0.0003$
\\
& \nameref{strat3} & \cellcolor{\bestres}$2.6998 \pm 0.0567$ & \cellcolor{\bestres}$1.3535 \pm 0.0209$ & \cellcolor{\bestres}$0.9807 \pm 0.0003$ & \cellcolor{\bestres}$0.9810 \pm 0.0002$
\\
\bottomrule
\end{tabular}
}
\end{center}
\end{table*}

\textbf{Results.}
Since the system is stable, we can compare all strategies without encountering the instability issue noted in \ref{item:D1}. The results are summarized in \autoref{tab:robot}. For $\sigma=10$, the excitation signal has less influence on the system input $\trueInMat$ than the feedback controller, resulting in a less informative $\trueInMat$. As direct methods (\nameref{strat1} and \nameref{strat2}) rely on $\trueInMat$, they perform poorly under such conditions. In contrast, \nameref{strat3} remains effective, as it identifies the system based on $\excMat$ instead of $\trueInMat$. It can be seen from \autoref{tab:robot} that \nameref{strat3} achieves lower OL and CL MSE and higher R\textsuperscript{2} scores.

For $\sigma=50$, all strategies are expected to improve due to increased excitation, which is confirmed by the increase in OL and CL R\textsuperscript{2}. Note that MSE is not suitable for assessing improvement in this case, as it is sensitive to the scale of the output signal, which increases significantly between experiments. While stronger excitation benefits all methods, \nameref{strat3} consistently outperforms the baselines across all evaluation criteria, further demonstrating its effectiveness.

\section{Conclusions and future work}
This paper presents a novel computationally tractable framework for closed-loop identification of nonlinear dynamical systems using a complete parameterization of systems stabilized by a given controller. The proposed method addresses key challenges, including instability risks and computational complexity. We theoretically show asymptotic consistency in the linear setting, aligning with established results. Numerical experiments demonstrate that our method outperforms existing work by accurately capturing system instabilities and achieving superior prediction performance.

Future work includes conducting a non-asymptotic statistical analysis in the spirit of \cite{Jones22nonasymp}, applying the identified model to adaptive control, and designing an iterative procedure that integrates system identification and control.

\balance
\bibliographystyle{IEEEtran}
\bibliography{root.bib}

\appendix
\section{Appendix}

\subsection{Proof of \autoref{theo:ici}}\label{app:proof_ici}
We first state a useful corollary from \cite{Ho20Corol}.
\begin{corollary}\label{coroll:inverse}
Consider a causal operator $\upOp$ such that $\upOpKnot := \upOp - \eyeOp$ is strictly causal, where $\eyeOp$ is the identity operator.
Then, the causal operator $\upOp^{-1}$ exists, and the sequence $\bMat = \upOp^{-1} \aMat$ satisfies, for $t = 0,1,\dots$:
\begin{align*}
    \bVec{t} = \aVec{t} - \upKnotFunc{t} (\bVec{t-1:0}).
\end{align*}
\end{corollary}

We start by proving \ref{item:T1}, meaning that for any $\noiseMat \in \lp^\dimOut$ and $\excMat \in \lp^\dimIn$, one has $\approxNoisyOutMat \in \lp^\dimOut$ and $\approxInMat \in \lp^\dimIn$.
    We denote the input to the operator $\approxOp$ by $\whatMat$. Then, from the right panel of \autoref{fig:true_CL_Ghat},
    \begin{align*}
        \whatMat = \approxInMat - \cont (\approxOutMat) = \excMat + \cont \bigl(\approxOp(\whatMat) + \noiseMat \bigr) - \cont \bigl(\approxOp(\whatMat) \bigr).
    \end{align*}
    We now upper bound its norm:
    \begin{align*}
        \Vert \whatMat \Vert_p &= \bigl\Vert \excMat + \cont \bigl(\approxOp(\whatMat) + \noiseMat \bigr) - \cont \bigl(\approxOp(\whatMat) \bigr) \bigr\Vert_p \\
        &\leq \bigl\Vert \excMat \bigr\Vert_p +\bigl \Vert \cont \bigl(\approxOp(\whatMat) + \noiseMat \bigr) - \cont \bigl(\approxOp(\whatMat) \bigr) \bigr\Vert_p \\
        &\leq \bigl\Vert \excMat \bigr\Vert_p + \gamma_K \bigl\Vert \approxOp(\whatMat) + \noiseMat - \approxOp(\whatMat) \bigr\Vert_p = \bigl\Vert \excMat \Vert_p + \gamma_K \Vert \noiseMat \bigr\Vert_p,
    \end{align*}
    where the second line is due to the triangle inequality and $\gamma_K>0$ is the i.f.g. of $\cont$.
    Since $\noiseMat \in \ell_p^\dimOut$ and $\excMat \in \ell_p^\dimIn$, we have $\whatMat \in \ell_p^\dimIn$. 
    Additionally, since $\approxOp \in \Lp$, $\approxNoisyOutMat = \approxOp(\whatMat) + \noiseMat \in \ell_p^\dimOut$. Consequently, since $\cont \in \Lp$, $\approxInMat = \cont (\approxNoisyOutMat) + \excMat \in \ell_p^\dimIn$.
    Therefore, the mapping from $(\excMat, \noiseMat)$ to $(\trueInMat, \trueNoisyOutMat)$ is an $\Lp$ operator.

Next, we prove \ref{item:T2}.
Given a strictly causal system $\approxSys$ stabilized by $\cont$, define $\approxOp = \approxSys(\eyeOp - \cont \approxSys)^{-1}$. Note that the inverse of $(\eyeOp - \cont \approxSys)$ exists due to \autoref{coroll:inverse}.
Additionally, set the initial condition of $\approxOp$ such that:
\begin{align}
    \approxOpFunc{0}(\varnothing - \contFunc{0}(\varnothing)) = \approxSysFunc{0}(\varnothing).\label{eq:Shatinit}
\end{align}
In~\cite{Desoer82GlobalPO}, it has been shown that $\approxOp \in \Lp^\textit{SC}$.

We apply the same input $\approxInMat$ to both $\approxSys$ and the closed-loop system formed by $\approxOp$ and $\cont$, as illustrated in the right panel of \autoref{fig:true_CL_Ghat}, and denote their respective outputs by $\approxOutMat{^{\textit{des}}}$ and $\approxOutMat$. Our goal is to prove that $\approxOutMat{^{\textit{des}}} = \approxOutMat$, or equivalently, $\approxOut{t}{^{\textit{des}}} = \approxOut{t}$ for all $t \geq 0$. At $t=0$, this is ensured by \eqref{eq:Shatinit}.

First, we compute the outputs. For $\approxSys$, we have $\approxOutMat{^{\textit{des}}}=\approxSys (\approxInMat)$.
From \eqref{eq:ici}, we obtain that:
\begin{align*}
    \approxOutMat &= \approxOp \bigl( \approxInMat - \cont (\approxOutMat) \bigr)
    = \approxSys \underbrace{(\eye - \cont \approxSys)^{-1}\bigl( \approxInMat - \cont (\approxOutMat) \bigr)}_{\bMat} = \approxSys \, \bMat.
\end{align*}
Converting the above into time equations gives:
\begin{align}
    \approxOut{t}{^{\textit{des}}}=\approxSysFunc{t} (\approxIn{t-1:0})
    \quad , \quad
    \approxOut{t} = \approxSysFunc{t}(\bVec{t-1:0}), \label{eq:proof_ici_timedomain}
\end{align}
where we use the strong causality of $\approxSys$ to truncate the operand at time $t-1$. 
Additionally, $\bVec{t}$ can be computed using the recursive algorithm of \autoref{coroll:inverse} as:
\begin{align}\label{eq:bt2}
    \bVec{t} = \approxIn{t} - \contFunc{t}(\approxOut{t:0}) + \contFunc{t} \bigl( \approxSysFunc{t:0}(\bVec{t-1:0})\bigr) \quad \forall t\geq 0.
\end{align}

From \eqref{eq:proof_ici_timedomain}, one has $\approxOut{t}{^{\textit{des}}} = \approxOut{t}$ for all $t \geq 0$ if $\approxIn{t} = \bVec{t}$ for all $t \geq 0$. By \eqref{eq:bt2}, this holds if $\contFunc{t}(\approxOut{t:0}) = \contFunc{t} \bigl(\approxSysFunc{t:0}(\bVec{t-1:0})\bigr)$ for all $t\geq 0$. Since $\approxOut{t} = \approxSysFunc{t:0}(\bVec{t-1:0})$ from \eqref{eq:proof_ici_timedomain}, the condition is satisfied, thus completing the proof.

\end{document}